# Mott Transition and Superconductivity in Quantum Spin Liquid Candidate NaYbSe$_2$*

Ya-Ting Jia (贾雅婷)[1,2,#], Chun-Sheng Gong (龚春生)[3,#], Yi-Xuan Liu (刘以轩)[3], Jian-Fa Zhao (赵建发)[1], Cheng Dong (董成)[4], Guang-Yang Dai (代光阳)[1], Xiao-Dong Li (李晓东)[5], He-Chang Lei (雷和畅)[3,**], Run-Ze Yu (于润泽)[1,**], Guang-Ming Zhang (张广铭)[6,7], and Chang-Qing Jin (靳常青)[1,2**]

[1]*Institute of Physics, Chinese Academy of Sciences, Beijing 100190, China*
[2]*University of Chinese Academy of Sciences, Beijing 100190, China*
[3]*Department of Physics and Beijing Key Laboratory of Opto-electronic Functional Materials & Micro-nano Devices, Renmin University of China, Beijing 100872, China*
[4]*Peking University Shenzhen graduate School, School of Advanced Materials, Shenzhen 518055, China*
[5]*Beijing Synchrotron Radiation Facility, Institute of High Energy Physics, Chinese Academy of Sciences, Beijing, 100049, China*
[6]*State Key Laboratory of Low-Dimensional Quantum Physics and Department of Physics, Tsinghua University, Beijing 100084, China*
[7]*Frontier Science Center for Quantum Information, Beijing 100084, China*

*Supported by the National Key R&D Program of China (Grants No. 2016YFA0300504, 2018YFE0202600 and 2018YFA0305701), the National Natural Science Foundation of China (Grants No. 11774423, 11822412), the Fundamental Research Funds for the Central Universities, and the Research Funds of Renmin University of China (RUC) (18XNLG14, 19XNLG17).
**Email: hlei@ruc.edu.cn; yurz@iphy.ac.cn; jin@iphy.ac.cn
#These authors contribute equally to this work.

**Abstract** The Mott transition is one of the fundamental issues in condensed matter physics, especially in the system with antiferromagnetic long-range order. However, such a transition is rare in quantum spin liquid (QSL) systems without long-range order. Here we report the experimental pressure-induced insulator to metal transition followed by the emergence of superconductivity in the QSL candidate NaYbSe$_2$ with a triangular lattice of 4$f$ Yb$^{3+}$ ions. Detail analysis of transport properties in metallic state shows an evolution from non-Fermi liquid to Fermi liquid behavior when approaching the vicinity of superconductivity. An irreversible structure phase transition occurs around 11 GPa is revealed by the X-ray diffraction. These results shed light on the Mott transition in the QSL systems.

Frustrated magnets are materials in which localized magnetic moments (spins), interact through competing exchange interactions that cannot be simultaneously satisfied. Magnetic frustration systems have attracted tremendous interests because they exhibit numerous exotic emergent phenomena[1]. One of important examples is quantum spin liquid (QSL) with a ground state of strong quantum fluctuations preventing the phase transition towards conventional magnetic order. It exhibits long-range quantum entanglement[1-5] and de-confined spinon excitations that may obey varied statistics rules of boson, fermion, or even anyon depending on the types of QSL[2]. More importantly, P. W. Anderson proposed that the superconductivity in copper oxide superconductors can evolve from spin liquid state[6]. Therefore, Mott transition and superconductivity emerging from the QSL are very interesting.

Mott transition is one of the important subjects in the physics of strongly correlated electrons especially in the system with antiferromagnetic long-range order[7]. However, the ground state of the Mott insulator in QSLs is not trivial and Mott transition between metallic and insulating spin-liquid phase is very rare[1]. For example, attempting to dope QSL candidate $ZnCu_3(OH)_6Cl_2$ indicated there was no observation for metallic or superconducting states even up to 0.6 electron per $Cu^{2+}$ and down to 1.8 K[8]. Theoretical analysis found the localization of electrons in various Zn-Cu hydroxyl halides and the formation of polaronic states with attendant lattice displacements and a dramatic narrowing of bandwidth upon electron addition[9]. Compare with charge carriers doping, pressure is an effective and clean approach to induce Mott transition. For example, pressure-induced (band-width-controlled) metal-insulator transition (MIT) with exotic quantum criticality and superconductivity appears in the vicinity of MIT with possible unconventional mechanisms in organic QSL candidates $\kappa$-$(ET)_2Cu_2(CN)_3$ and $\kappa$-$(ET)_2Ag_2(CN)_3$[10-16], in which the Hubbard $U$ is relative small. In contrast, although external pressure can modify magnetic properties of quasi-two-dimensional

(quasi-2D) inorganic frustrated systems significantly by way of tuning magnetic exchange interaction, crystal energy filed (CEF), dipole interaction[17], the insulating behavior still persists under high pressure, such as in α-RuCl$_3$ and α-Li$_2$IrO$_3$[18-20], possibly due to the large on-site Coulomb repulsion in these strong Mott insulators. Takagi *et al.* claimed the observation of the pressure induced insulating to metallic phase transition in QSL candidate Na$_4$Ir$_3$O$_8$[21], but there was no report on the details of experiment. Thus, the MIT in quasi-2D inorganic magnetic frustrated inorganic materials is still elusive.

Very recently, a novel material system NaLnCh$_2$ (Ln = rare earth, Ch = O, S, Se), especially NaYbCh$_2$, has been proposed to be a promising candidate to realize QSL state[22-27]. NaYbCh$_2$ has a perfect triangle lattice and similar CEF environment of 4$f$ Yb$^{3+}$ ions to YbMgGaO$_4$[28]. Thus NaYbS$_2$ is an effective spin-1/2 quantum magnet at low temperatures. It possesses the simplest structure and chemical formula among the known QSL candidates. In addition, the antiferromagnetic exchange interaction is significantly enhanced ($|\theta_{CW}| > 10$ K)[22-24], and the site disorder of elements is absent in NaYbCh$_2$ compared to YbMgGaO$_4$. Importantly, the long-range magnetic order or spin glass states are not observed in NaYbCh$_2$ when temperature down to 50 mK, generating an empirical frustration parameter $f = \theta_{CW}/T_N > 200$[22-25]. Continuous magnetic excitations are also observed in NaYbO$_2$ by inelastic neutron scattering show a gapless feature[25]. These results strongly suggest that NaYbCh$_2$ with strong spin-orbital coupling (SOC) could host a QSL state as the ground state. Moreover, the absorption spectra indicate that the charge gaps are roughly 4.5 eV, 2.7 eV and 1.9 eV for NaYbO$_2$, NaYbS$_2$ and NaYbSe$_2$, respectively. The variable and small charge gaps may allow the system to access a MIT by applying pressures.

In this work, we study the pressure effects on structural and transport properties of NaYbSe$_2$ up to 125.9 GPa. It is found that there is a structural transition from the *R*-3*m*H to *P*-3*m*1 at around 11 GPa. A MIT appears at about 58.9 GPa, accompanying with the giant change of resistivity by eight orders of magnitude. A superconducting transition emerges when pressure is higher than 103.4 GPa.

Single crystals of NaYbSe$_2$ were grown by NaCl flux method. The mixture of Yb powder, Se grain and NaCl grain with the molar ratio of 2 : 3 : 10 was loaded into a silica tube and then sealed under Ar atmosphere (∼ 0.2 atm). The sealed tube was heated to 1173 - 1223 K for 24 h and kept at that temperature for 150 h. Then it was cooled down to room temperature naturally. Finally, the single crystals were separated by washing NaCl flux with distilled water for several times. NaYbSe$_2$ powders were synthesized by solid state reaction method as described in elsewhere[22].

The electronic transport properties of NaYbSe$_2$ under high pressure and low temperatures were investigated via Van der Pauw method in a diamond anvil cell (DAC) made of CuBe alloy as described in references[29-33]. Pressure was generated by a pair of diamonds with a 100 $\mu$m diameter culet. A gasket made of T301 stainless steel was pressed from a thickness of 250 $\mu$m to 20 $\mu$m, and drilled a center hole with a diameter of 300 $\mu$m. Fine cubic boron nitride (cBN) powder was used to cover the gasket to protect the electrode leads insulated from the metallic gasket. The electrodes were slim Au wires with a diameter of 18$\mu$m. A 50$\mu$m-diameter center hole in the insulating layer was used as the sample chamber. The dimension of the sample was about 45 $\mu$m × 45 $\mu$m × 5 $\mu$m, and NaCl powder was as the pressure transmitting medium. The pressure was measured via the ruby fluorescence method at room temperature before and after each cooling[34]. The diamond anvil cell was placed inside a MagLab system to perform the experiments. The temperature was automatically controlled by a program of the MagLab system. A thermometer was mounted near the diamond in the diamond anvil cell to monitor the exact sample temperature.

In-situ high pressure angle-dispersive X-ray diffraction (ADXRD) experiments were performed using a symmetric Mao Bell DAC at Beijing Synchrotron Radiation Facility. The wavelength is 0.6199 Å. The sample in DAC is fine NaYbSe$_2$ powder and a tiny ruby chip was regarded as the pressure marker. The two dimensional image plate patterns obtained were converted to one-dimensional 2$\theta$ versus intensity data using the Fit2d software package[35].

Fig. 1(a) shows the evolution of in-plane resistance $R(T)$ as a function of temperature for NaYbSe$_2$ single crystal at various pressures below 49.6 GPa. Before 8 GPa, we can't measure the $R(T)$ curve via four-probe method because of high resistance of NaYbSe$_2$. All of $R(T)$ curves show an insulating behavior that increases with decreasing temperature, but the absolute values of $R(T)$ at room temperature decrease with increasing pressure dramatically by more than eight orders of magnitude. The transport behavior can be fitted well by using a variable range hopping model $R(T) = R_0\exp(T_0/T)^{1/(d+1)}$, where $T_0$ is the characteristic temperature and $d = 2$ for the two dimensional system. This formula is usually used to descript the weak conducting behavior of Mott insulator. The inset of Fig. 1(b) shows the fitting result of $R(T)$ at $P = 42$ GPa and the fitted $T_0$ is shown in the main panel of Fig. 1(b). It can be seen that the $T_0$ decreases gradually with pressure, i.e., the hopping barrier or band gap decreases under pressure. When increasing pressure further ($P = 58.9$ GPa), a metallic state is observed (Fig. 1(c)), but with a minimum around 55 K (Defined as $T_{min}$). After carefully checking the resistance below $T_{min}$, we find that it obeys a logarithmic temperature dependence (Inset of Fig. 1(c)), which may be explained by either weak localization effect originated from the presence of disorder potentials or incoherent Kondo effect due to the presence of quantum magnetic impurities[36]. Considering that the electric resistivity of the samples in the paramagnetic insulating phase is fitted to the variable range hopping model, we tend to interpret the resistivity up-turn at low temperatures as the weak localization effect. This minimum shifts progressively to lower temperatures under higher pressure, similar to that observed in Yb-series heavy-fermion compounds[37]. Then a complete metallic behavior in the whole temperature range (2 - 300 K) is achieved at $P_c = 74.8$ GPa. The piezochromism of the NaYbSe$_2$ single crystal also reflects the narrowing of hopping barrier or band gap (Fig. 1(d)). The color of sample at low pressure is brown-red, which is coincided with the result reported in previous work[22]. The sample becomes much darker with the increase of pressure, and only a small amount of red color can be seen on the edge at $P = 37.5$ GPa, indicating the gradually decrease of hopping barrier or band gap[38]. Finally, the sample becomes completely dark when $P$ is above 42.2 GPa, suggesting that the band gap is very small and the light can't go through sample.

Under higher pressure, the values of $R(T)$ continuously decrease and metallic behavior is shown as in Fig. 2(a). Surprisingly, when $P = 103.4$ GPa, a sudden drop at

$T \sim 8$ K appears on the $R(T)$ curve, suggesting the emergence of superconductivity (Fig. 2(b)). Although the superconducting transition temperature $T_c$ is almost unchanged with pressure, the resistance drop becomes more explicit (inset of Fig. 2(b)). In addition, with increasing magnetic field along the $c$ axis, the resistance drop at $P = 125.9$ GPa is suppressed gradually (Fig. 2(c)), confirming the drop on $R(T)$ curve at around 8 K originates from the superconducting transition indeed. Our result is not related to any results of pressure induced superconductivity for the single element or binary of Na-Yb-Se. So we can also rule out the possibility from impurities. It is noted that the $R(T)$ does not vanish at 125.9 GPa, which is the reachable maximum pressure at present experiment condition. This may also be induced by the imperfect sample quality or pressure inhomogeneity[39].

In order to analyze the metallic state under pressure, we have applied the power-law fitting $R(T) = R_0 + AT^n$ to the resistance of NaYbSe$_2$ for $P \geq 74.8$ GPa. The inset of Fig. 2(d) shows the fitting result of $R(T)$ between 8 K and 100 K at 74.8 GPa. It can be seen that the power-law formula fits the experimental data well in the temperature range between 8 K and 100 K. The pressure dependence of the exponent $n$ is plotted in Fig. 2(d). At the boundary of MIT, the metallic state of NaYbSe$_2$ clearly shows a non-Fermi-liquid (NFL) behavior with $n \sim 1$. With increasing pressure, the $n$ increases gradually and finally gets close to 2 ($n = 2.1(1)$ at $P = 125.9$ GPa), implying the Fermi-liquid (FL) behavior in NaYbSe$_2$. Therefore, there is a crossover from NFL to FL behavior with increase of pressure as approaching the boundary of superconductivity. Usually, it is argued that the values of $n<2$ are caused by quantum critical fluctuations. This phenomena have been observed frequently in heavy-fermion systems where magnetic orders are suppressed by doping, magnetic field or pressure[40-43]. But it should be mentioned that the NFL behavior observed in NaYbSe$_2$ is not confined to the vicinity of $P_c$, but extends to much higher pressures. More importantly, different from the situation in $\kappa$-(ET)$_2$Cu$_2$(CN)$_3$ and $\kappa$-(ET)$_2$Ag$_2$(CN)$_3$ that superconductivity usually emerges in the quantum-critical-fluctuation region near the end point of a MIT[10, 11, 16, 43], the superconductivity in NaYbSe$_2$ seems to appear when $P$ is away from the $P_c$ of

MIT and the metallic state exhibits a FL behavior. Such interesting features need to be studied at even higher pressure in order to check whether the value of $n$ is still near 2 in the bulk superconducting region in the future.

Fig. 3(a) shows the synchrotron X-ray diffraction patterns under various pressure. At ambient pressure, the pattern can be indexed very well with a space group $R$-$3m$H as reported in the previous work (named phase I)[22]. The peaks shift to the high angle with the increasing of pressure, indicating the shrink of the lattice parameters. The additional peaks around 18.6 degree appear (See arrow in Fig. 3(a)) at around 12 GPa. Finally, it completely transfers into a new phase (named phase II) at around 19.4 GPa. The phase II keeps stable with pressure up to 43.5 GPa. When the pressure is released, the high pressure phase can be kept, indicating that the phase transition is irreversible. Carefully checking the X-ray diffraction patterns under pressure, it can be seen that almost all the patterns at low pressure phase persist to the high pressure phase. For example, the peak at around 5.1 degree at ambient pressure (the peak of (003) in phase I), which describes the framework of this hexagonal structure along $c$ axis, keeps alive up to the highest pressure in the experiment. The framework of the low pressure phase is kept at high pressure, and there should be the relative change for the atoms during the phase transition.

After carefully checking the pattern at high pressure, we find the high pressure phase pattern can be reproduced with a structure with space group $P$-$3m$1. The details of crystal structure refinement and data can be found in Fig. S1 and Table S1. The crystal structures viewed along $c$ axis and $ab$ plane for phase I and II are shown in Figs. 3(b) and (c). Both structures show the same hexagonal lattice structure along $c$ axis (See Fig. 3(b)). While the Na and Se layers are pushed to approach to each other under high pressure in the $ab$ plane, compared with the lower pressure phase. In the high pressure phase, two different types of YbSe6 octahedra stack separately along $c$ axis. Their relative movement between Na and Se layers induces the elongation YbSe6 octahedra along $c$ axis, but it does not change the perfect triangular network of Yb ions. Thus, the features of in-plane magnetic frustration should be still intact for the high pressure phase.

Fig. 4 exhibits a pressure-temperature phase diagram of NaYbSe$_2$ single crystal. As seen clearly, the application of high pressure reduces the resistance continuously and a metallic state at 58.9 GPa with a weak localization at low temperatures. Then superconductivity appears at 103.4 GPa. There is a transition from non-Fermi liquid to Fermi liquid state at metallic areas. Meanwhile, according to the results of ADXRD, a structural phase transition occurs from low pressure phase *R*-3*m*H to a high pressure phase with *P*-3*m*1. The pressure-turned Mott transition and superconductivity is observed at the second phase. Such *P-T* phase is different from that observed in organic QSL compound $\kappa$-(ET)$_2$Cu$_2$(CN)$_3$[16], in which superconductivity emerges from the quantum-critical-fluctuation region near the end point of a MIT, indicating there may be two different mechanisms for the electronic states evolution in these two systems. Meanwhile we noticed that Zhang *et al.* conducted a rather similar study on NaYbSe$_2$ at relatively low pressure region and with very different phase diagrams[44]. According our experimental result, this discrepancy may originate from the inhomogeneity of the samples. It should be mentioned that compounds containing *f*-electron elements usually display a wealth of superconducting phases, such as heavy fermion superconductivity, which are prime candidates for unconventional superconductivity with complex order parameter pairing symmetries[45]. As to NaYbSe$_2$, the competition/collaboration of the Kondo spin exchange interaction, magnetic frustration and superconductivity may bring a plenty of interesting physical properties[46]. Undoubtedly, further experimental and theoretical investigations are needed in order to identify whether it is an unconventional superconductor as proposed in the triangular lattice Na-doped cobaltates[47-48].

In summary, we have discovered pressure-induced MIT at 58.9 GPa and superconductivity appearing at much higher pressure away from MIT (*P* = 103.4 GPa) in QSL candidates NaYbSe$_2$, accompanying with a structural phase transition around 11 GPa from the *R*-3*m*H to *P*-3*m*1. The low temperature *R*(*T*) in the metallic state exhibits a crossover from NFL to FL behavior. These observations open up a promising way to study the features of MIT and the interplay between spin and charge degrees of freedom in the QSL system with strong SOC. Furthermore, a large family of NaLnCh$_2$ compounds provides a novel platform to investigate the effects of 4*f* and chalcogen ions on the possible MIT and superconductivity in magnetic frustration systems.

**Figure captions:**

**Figure 1.** Insulator to metal transition in NaYbSe$_2$. (a) Temperature dependence of resistance $R(T)$ for NaYbSe$_2$ single crystal in insulating state with pressure up to 49.6 GPa. (b) Pressure dependence of fitted $T_0$ using variable range hopping model $R(T) = R_0\exp(T_0/T)^{1/(d+1)}$ ($d = 2$ for the two dimensional property). The inset shows the fitting result at 42 GPa. (c) Temperature dependent resistance $R(T)$ for NaYbSe$_2$ single crystal between 58.9 GPa and 70.4 GPa. The arrows denote the minimums of $R(T)$ (The temperature at minimum of $R$ (T) is noted as $T_{min}$). The inset shows the resistance vs log$T$ data at 66.1 GPa, showing a linear behavior below $T_{min}$ (See black line). (d) Piezochromism of the NaYbSe$_2$ single crystal at various pressures, demonstrating the decrease of band gap with increasing pressure.

**Figure 2.** Pressure induced superconductivity in NaYbSe$_2$. (a, b) Temperature dependence of $R(T)$ for NaYbSe$_2$ single crystal in metallic state between 74.8 GPa and 98.9 GPa, and between 103.4 GPa to 125.9 GPa, respectively. The drop of $R(T)$ in b indicates the appearance of superconductivity. The inset of b shows the enlarged parts of normalized $R(T)/R(20\ K)$ at various pressures. (c) Temperature dependence of $R(T)$ at $P = 125.9$ GPa at different magnetic fields, confirming that the drop of $R(T)$ curve around 8 K is a superconducting transition. (d) Pressure dependence of fitted $n$ in metallic state using the formula $R(T) = R_0 + AT^n$. Inset: experimental and fitting results of $R(T)$ below 100 K at $P = 74.8$ GPa.

**Figure 3.** Structural transition under pressure. (a) Synchrotron X-ray diffraction patterns of NaYbSe$_2$ at selected pressures. Additional peaks of new phases are marked by arrow. (b) and (c) Crystal structure of NaYbSe$_2$ Phase I and II viewed along $c$ axis and $ab$ plane, respectively.

**Figure 4.** Pressure-temperature ($P$-$T$) phase diagram of NaYbSe$_2$ single crystal. "WL", "SC", "QSL", "FL" and "NFL" are abbreviations for weak localization, superconducting, quantum spin liquid, Fermi liquid and non-Fermi-liquid, respectively. Blue and brown dots denote the phase boundaries of WL-metal and metal-superconductivity.

**Figure 1**

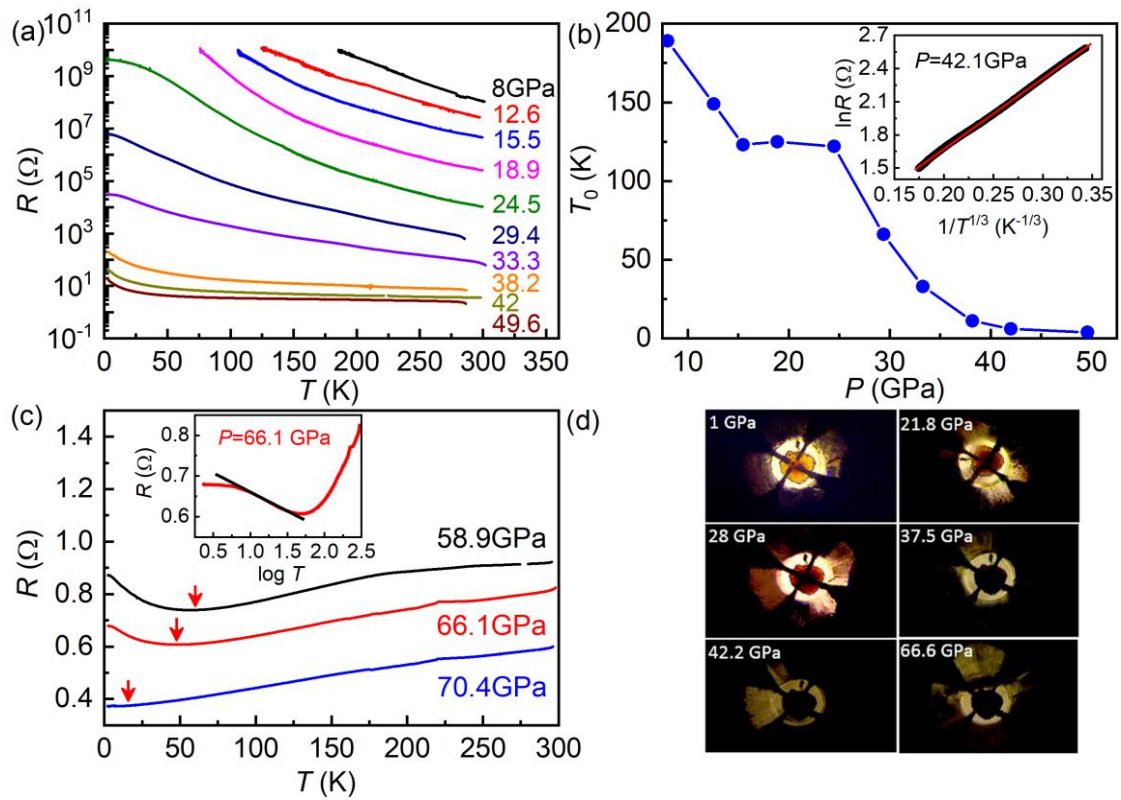

**Figure 2**

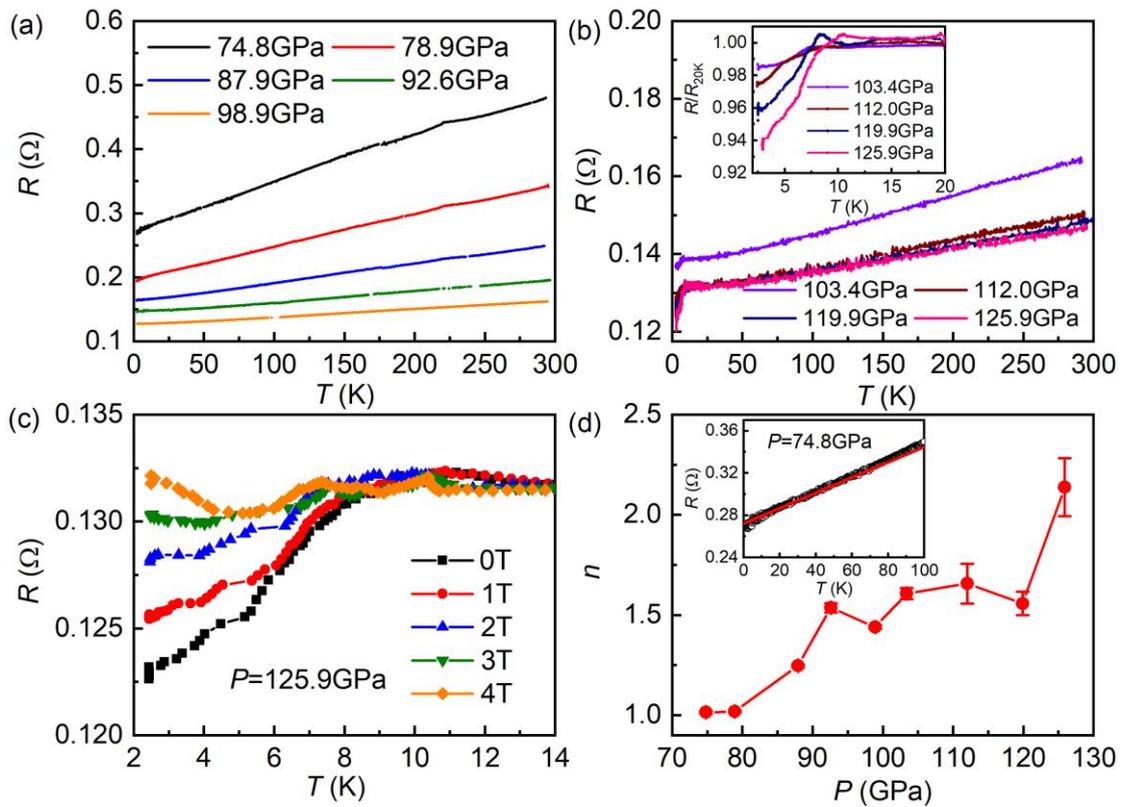

**Figure 3**

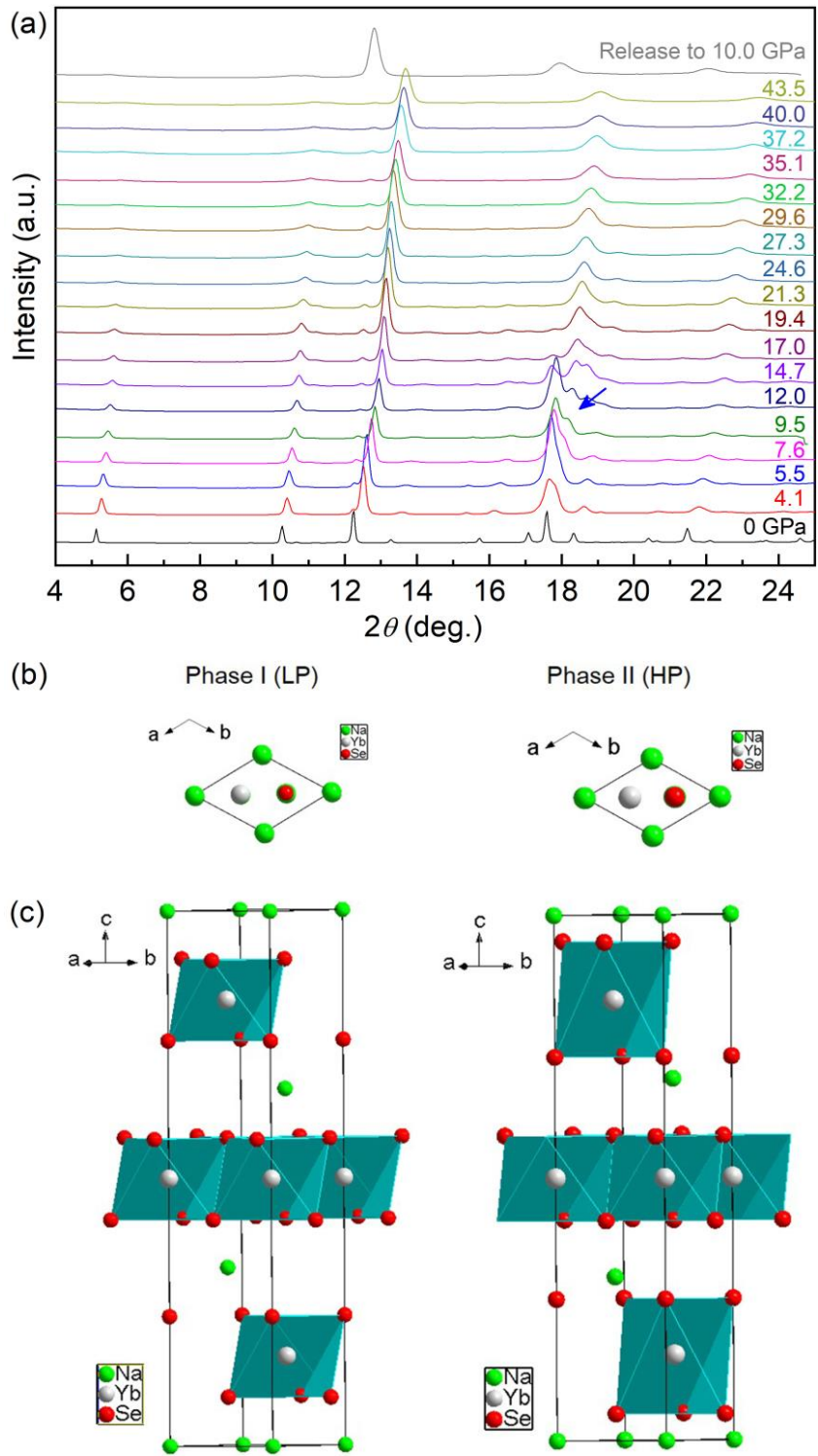

**Figure 4**

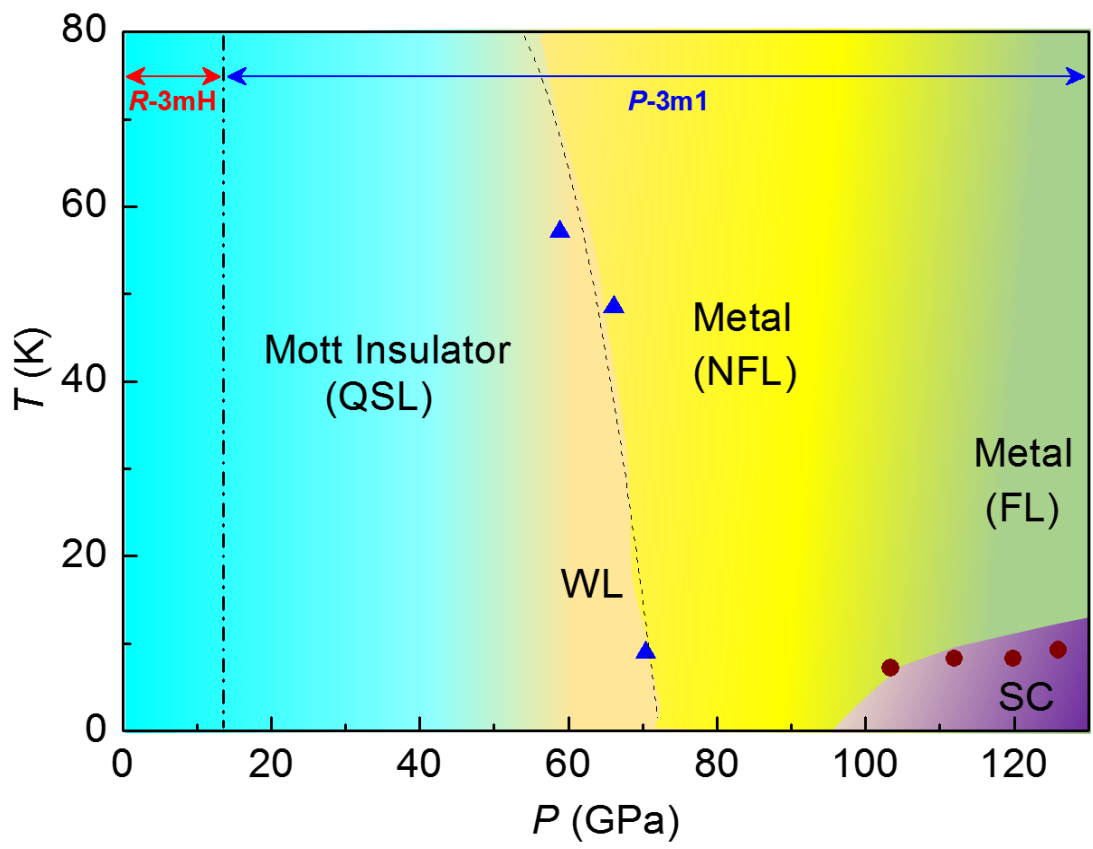

# Supplementary Information
# Mott Transition and Superconductivity in Quantum Spin Liquid Candidate NaYbSe$_2$

Ya-Ting Jia (贾雅婷)[1,2,#], Chun-Sheng Gong (龚春生)[3,#], Yi-Xuan Liu (刘以轩)[3], Jian-Fa Zhao (赵建发)[1], Cheng Dong (董成)[4], Guang-Yang Dai (代光阳)[1], Xiao-Dong Li (李晓东)[5], He-Chang Lei (雷和畅)[3,**], Run-Ze Yu (于润泽)[1,**], Guang-Ming Zhang (张广铭)[6,7], and Chang-Qing Jin (靳常青)[1,2**]

[1]*Institute of Physics, Chinese Academy of Sciences, Beijing 100190, China*
[2]*University of Chinese Academy of Sciences, Beijing 100190, China*
[3]*Department of Physics and Beijing Key Laboratory of Opto-electronic Functional Materials & Micro-nano Devices, Renmin University of China, Beijing 100872, China*
[4]*Peking University Shenzhen graduate School, School of Advanced Materials, Shenzhen 518055, China*
[5]*Beijing Synchrotron Radiation Facility, Institute of High Energy Physics, Chinese Academy of Sciences, Beijing, 100049, China*
[6]*State Key Laboratory of Low-Dimensional Quantum Physics and Department of Physics, Tsinghua University, Beijing 100084, China*
[7]*Frontier Science Center for Quantum Information, Beijing 100084, China*

**Email: hlei@ruc.edu.cn; yurz@iphy.ac.cn; jin@iphy.ac.cn

[#]These authors contribute equally to this work.

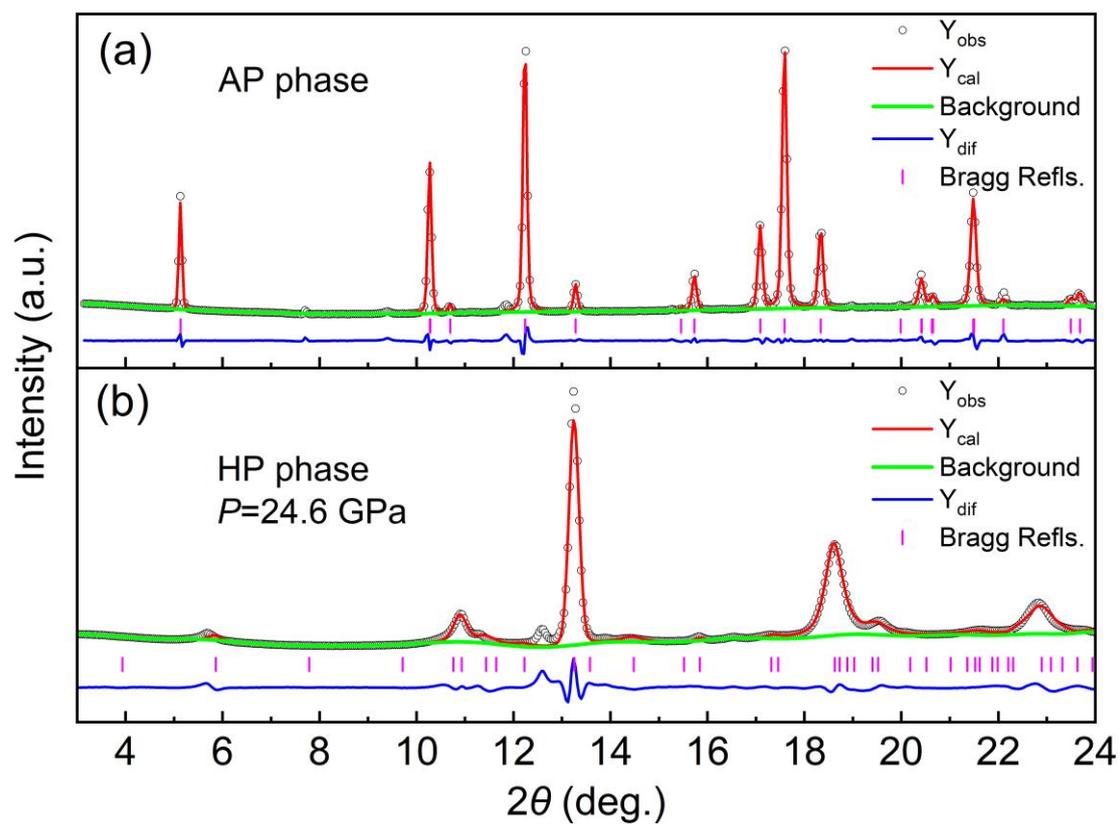

**FIG S1.** Rietveld refinement of XRD patterns. (a) low pressure phase (phase I, 0 GPa, *R*-3*m*H) and (b) high pressure phase (phase II, 24.6 GPa, *P*-3*m*1). Experimental pattern, calculated pattern and their difference shows as black cross, red curve and blue curve, respectively. The pink bars represent the position of Bragg reflections. The wavelength of the X-ray is 0.6199 Å.

**Table S1.** Refined crystal structural parameters for the low pressure phase (Phase I) and high pressures phase (phase II, 24.6 GPa) data.

|  | Phase I | Phase II |
|---|---|---|
| Space group | $R$-$3m$H | $P$-$3m1$ |
| $a$ (Å) | 4.0536(8) | 3.8487(5) |
| $c$ (Å) | 20.7462(4) | 18.4729(8) |
| $V$ (Å$^3$) | 295.23(6) | 236.97(7) |
| Calc. Density (g/cm$^3$) | 5.97(2) | 7.44(1) |
| Na1 | (0, 0, 0) | (0, 0, 0) |
| Na2 |  | (1/3, 2/3, -0.3117) |
| Yb1 | (0, 0, 0.5) | (0, 0, 0.5) |
| Yb2 |  | (1/3, 2/3, 0.1626) |
| Se1 | (0, 0, 0.2444) | (0, 0, 0.2697) |
| Se2 |  | (1/3, 2/3, -0.0508) |
| Se3 |  | (1/3, 2/3, 0.4195) |
| $U_{iso}$Na1 | 0.0151(3) | 0.0467(4) |
| $U_{iso}$Na2 |  | 0.0126(7) |
| $U_{iso}$Yb1 | 0.0016(8) | 0.0483(0) |
| $U_{iso}$Yb2 |  | 0.0142(6) |
| $U_{iso}$Se1 | 0.0021(2) | 0.0596(9) |
| $U_{iso}$Se2 |  | 0.0005(1) |
| $U_{iso}$Se3 |  | 0.0152(5) |
| $R_{wp}$ (%) | 3.35 | 4.82 |
| $R_p$ (%) | 2.02 | 2.73 |